# Evaluation of Italian astronomical production: 2010-2012

*Raffaele Gratton*

*INAF-Osservatorio Astronomico di Padova*

## Introduction

The purpose of this document is to present a few statistics about the role of Italian astronomy, focusing on the production by INAF (Istituto Nazionale di Astrofisica). Data are presented but very few comments are given. We did not use lists of members of different scientific INAF macro-areas[1] because a few trials showed that they are largely incomplete and result in an underestimate of the Italian astronomical production by more than a factor of two. The structure of this document is as follows: first I give some detail about the methods used; I then present data about the role of Italy in astronomical research worldwide; finally, I give some statistics about the h-factor of astronomers that are members of scientific INAF macro-areas (both INAF staff and associate).

## Methods

The basic method was queries to ADS[2] by affiliation; they were done within one week in early March, 2013. When not specified, all queries only refer to 2010-2012. A comparison was made with analogous statistics made for the period 2005-2007.

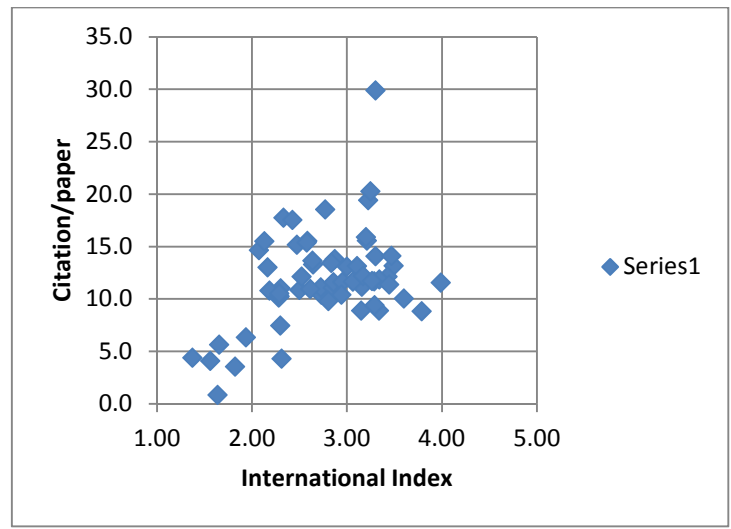

Figure 1. Correlation between internationalization index and number of citations per paper

---

[1] INAF scientific activity is organized in five Macro-areas: 1. Galaxies and Cosmology; 2. Stars, stellar populations and interstellar matter; 3. The Sun and the solar system; 4. High energy astrophysics and relativity; 5. Technology.
[2] http://adsabs.harvard.edu/mighty_search.html

When not specified, all queries are for citations. Many papers have authors from different countries. We define the internationalization index as the ratio:

int= Σ $cit_i$ / cit,

where $cit_i$ and cit are the number of citations for country i and for world, respectively. This internationalization index raised rapidly with time: it was 1.71 for 2001-03, 1.91 for 2004-06, 2.27 for 2007-09, and 2.72 for 2010-12. Neglecting it results in overestimating the contributions from each country and in spurious trends with time.

For a given keyword, the internationalization index is well correlated with the number of citations per paper (see Figure 1).

The impact of a country is then defined as:

Impact = average { $(cit_i/Σ cit_i)$, $(cit_i/Σ cit_i)^{vint}$}/ Σ average { $(cit_i/Σ cit_i)$, $(cit_i/Σ cit_i)^{vint}$}

Denominator is used to normalize to the world production. The second term in the averages is used to correct for the nation size effect (authors in small countries are more likely to collaborate with those of other countries than authors in big countries; the effect is expected to be larger in more internationalized topics). The formula is empirical; we deem it adequate because it produces near constant values over time for the two countries with the largest impact (United States and Germany), in front of a strong rise of the internationalization factor. Comparing with other formulas, we deem that the uncertainty due to the use of this particular functional form is less than 10% for Italy. It has no impact on rankings.

Special care was devoted to select the most appropriate journals and keywords; this is crucial in particular for Macro-areas 3 and 5; for these areas, contamination of keywords may produce misleading results. The following journals were considered:

- Macro-area 1, 2, and 4: ARA&A, AJ, AstRv, AN, A&A, A&AS, A&ARv, ApJ, ApJL, ApJS, Ap&SS, MNRAS, Natur, NewA, NewAR, PhRvD, PASJ, PASP, Science, GReGr, AIP, JCAP, APh, SPIE
- Macro-area 3: all refereed papers in Astrophysics and Physics
- Macro-area 5: SPIE

Notes for individual keywords are given in the Appendix.

# Impact of Italy on astronomical research

## General

### Time series of impact
Table 1 lists the top ten countries according to the average impact parameter. We give data for the four most recent three-years periods. We defined:

Average = [(2001-2003)+(2004-2006)+(2007-2009)+(2010-2012)]/4

Trend=1-{[(2007-2009)+(2010-2012)]- [(2001-2003)+(2004-2006)]}/[(2001-2003)+(2004-2006)]

Italy was fifth according to the impact parameter for each of these time periods. However, the trend of the impact with time is slightly negative though less than for United Kingdom, Japan, Netherland and Australia. There is no important trend for United States, Germany and France. Canada, Spain, and China (beyond the 10[th] position) are rapidly rising.

Table 1. Trends of the impact of different countries on world astronomy over the last 12 years

| Rank | Country | 2001-2003 | 2004-2006 | 2007-2009 | 2010-12 | Average | trend |
|---|---|---|---|---|---|---|---|
| 1 | United States | 0.354 | 0.342 | 0.346 | 0.346 | 0.346 | -0.004 |
| 2 | Germany | 0.110 | 0.118 | 0.114 | 0.114 | 0.114 | +0.002 |
| 3 | United Kingdom | 0.093 | 0.092 | 0.088 | 0.080 | 0.090 | -0.086 |
| 4 | France | 0.065 | 0.068 | 0.067 | 0.068 | 0.067 | +0.012 |
| *5* | *Italy* | *0.065* | *0.062* | *0.061* | *0.059* | *0.062* | *-0.060* |
| 6 | Canada | 0.039 | 0.033 | 0.046 | 0.050 | 0.041 | +0.329 |
| 7 | Japan | 0.037 | 0.036 | 0.033 | 0.030 | 0.035 | -0.121 |
| 8 | Spain | 0.030 | 0.033 | 0.035 | 0.045 | 0.034 | +0.284 |
| 9 | Netherland | 0.035 | 0.030 | 0.024 | 0.028 | 0.029 | -0.203 |
| 10 | Australia | 0.027 | 0.022 | 0.021 | 0.021 | 0.023 | -0.136 |

Table 2. Nations ranked by first authors of papers among the 200 most cited of each year in the period 2008-10

| Rank | Country | 1[st] author among 200 most cited each year (2008-10) | |
|---|---|---|---|
| | | total | fraction |
| 1 | United States | 297 | 0.495 |
| 2 | Germany | 69 | 0.115 |
| 3 | United Kingdom | 61 | 0.102 |
| *4* | *Italy* | *39* | *0.065* |
| 5 | France | 34 | 0.057 |
| 6 | Switzerland | 21 | 0.035 |
| 7 | Canada | 17 | 0.028 |
| 8 | Netherland | 15 | 0.025 |
| 9 | Australia | 9 | 0.015 |
| 10 | Japan | 6 | 0.010 |
| | Spain | 6 | 0.010 |

### First authors among the 200 most cited papers of each year (2008-10)

An alternative way to define the impact of a country is to consider the presence of each country among the 3x200=600 most cited papers of each year during the period 2008-2010. This is slightly less than the 1% most cited papers of every year on the main journals; it can be considered the elite of the world astronomical production. We considered here only the affiliation of the first author.

Table 2 lists the top ten countries according to this parameter. Figure 2 compares these two impact parameters. The correlation is very good (the linear correlation coefficient is r=0.99, which has an extremely high level of confidence); the leading role of United States in astronomy is underlined by the very high fraction of papers whose first author has an US affiliation among the 200 most cited papers of each year[3]. Italy is fourth in this special ranking.

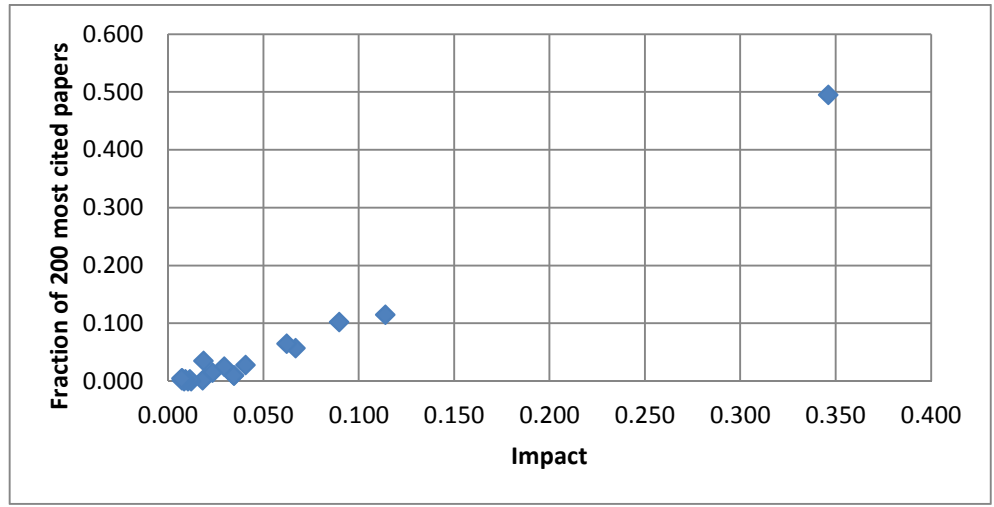

**Figure 2. Comparison between impact parameter and the fraction of papers within the 200 most cited (period 2008-2010).**

## More extensive time series

Hearnshaw (2006) obtained long term statistics using an approach quite similar to that considered here for the period 1976-2005, but limited to the number of papers (rather than on citations). From his data, it is possible to obtain the time series given in Table 3. I added to data by Hearnshaw the number of Italian papers within the 200 most cited for each year (since we considered 5 years bins, there are 1000 such paper for every time bin).

**Table 3. Long term trends for Italian astronomy in the world**

| Year | Ranking | Italian Papers | World Papers | Fraction | Italian papers among 200 most cited each year (total over a period of 5 years) |
|---|---|---|---|---|---|
| 1976-1980 | 6 | 1228 | 33005 | 0.037 | 13 |
| 1981-1985 | 6 | 2191 | 41114 | 0.053 | 20 |
| 1986-1990 | 6 | 2513 | 45203 | 0.056 | 25 |
| 1991-1995 | 7 | 2377 | 43015 | 0.055 | 31 |
| 1996-2000 | 5 | 3474 | 58388 | 0.059 | 49 |
| 2001-2005 | 5 | 7700 | 123837 | 0.062 | 52 |

These numbers show a clear growth of the incidence of Italian astronomy in the world over the thirty years 1976-2005. This contrasts with the slight decline of the last ten years mentioned above.

## Comparison with Gross Domestic Product

It is interesting to compare the astronomical production with the Gross Domestic Product (GDP) as listed by United Nations for 2011[4]. Table 4 lists the top ten countries according to GDP, with their impact on astronomy.

---
[3] The correlation between the two impact parameters is actually not linear, being better represented by a 1.5 power. This is not surprising: the "elite" is more concentrated in the leading countries than the "mass" production.

We may define the ratio between the fraction of the GDP and of the astronomical production within the 30 countries with the largest astronomical production. This is listed in the last column of this table.

We then listed in Table 5 the fifteen countries best ranked among this ratio. This is a measure of how much they produce in astronomy, normalized to their richness. Italy is sixth according to this ranking, behind United Kingdom, Netherland and Germany, similar to Switzerland, and in front of France, Canada, United States, Japan, and many other countries.

Table 4 First ten nations for GDP (in millions of US Dollars) and their impact in astronomy

| Rank | Country | GDP(2011) | Impact astronomy | Ratio |
|---|---|---|---|---|
| 1 | United States | 14991 | 0.346 | 1.38 |
| 2 | China | 7204 | 0.011 | 0.09 |
| 3 | Japan | 5870 | 0.035 | 0.35 |
| 4 | Germany | 3604 | 0.114 | 1.89 |
| 5 | France | 2776 | 0.067 | 1.44 |
| 6 | Brazil | 2477 | 0.006 | 0.14 |
| 7 | United Kingdom | 2429 | 0.090 | 2.21 |
| *8* | *Italy* | *2196* | *0.062* | *1.69* |
| 9 | India | 1898 | 0.007 | 0.23 |
| 10 | Russia | 1858 | 0.010 | 0.34 |

Table 5 First fifteen nations for ratio between impact in astronomy and GDP

| Rank | Country | Impact in astronomy/GDP |
|---|---|---|
| 1 | Chile | 4.33 |
| 2 | United Kingdom | 2.19 |
| 3 | Netherland | 2.09 |
| 4 | Germany | 1.88 |
| 5 | Israel | 1.77 |
| *6* | *Italy* | *1.68* |
| 7 | Switzerland | 1.67 |
| 8 | France | 1.43 |
| 9 | Canada | 1.39 |
| 10 | Spain | 1.38 |
| 11 | United States | 1.37 |
| 12 | Denmark | 1.25 |
| 13 | Portugal | 1.24 |
| 14 | South Africa | 1.23 |
| 15 | Sweden | 1.00 |

## Comparison to ESO countries

Over the whole period of 12 years, the total impact of the ESO countries to astronomy was 0.470 (rising from 0.462 in 2001-03 to 0.475 in 2010-12); the Italian fraction of this total is 0.132, which is very similar to the GDP

---
[4] http://en.wikipedia.org/wiki/List_of_countries_by_GDP_%28nominal%29

fraction (0.131). However, there is a slight negative trend (0.141, 0.130, 0.131, 0.123 for 2001-03, 2004-06, 2007-09, 2010-12, respectively).

## Impact by field

Table 6 lists average ranking over different keywords related to the specific Macro-areas, weighted according to the total number of citations worldwide, as well as the average impact parameter evaluated with the same weights. INAF impact is evaluated by considering the relative contributions of INAF, other research institutes, and Universities to the Italian astronomical production.

**Table 6. Impact of different macroareas on the respective fields of competence (2010-12 and for reference 2005-2007)**

| Area | | 2010-2012 | | | | 2005-2007 | | | |
|---|---|---|---|---|---|---|---|---|---|
| | | Average Ranking | Impact | | | Average Ranking | Impact | | |
| | | | Italy/World | INAF/Italy | INAF/World | | Italy/World | INAF/Italy | INAF/World |
| | General | 5.1 | 0.058 | 0.59 | 0.034 | 5.1 | 0.061 | 0.59 | 0.036 |
| MA1 | Galaxies | 4.5 | 0.064 | 0.65 | 0.042 | 4.7 | 0.062 | 0.62 | 0.039 |
| | Cosmology | 5.5 | 0.049 | 0.49 | 0.024 | 5.3 | 0.062 | 0.45 | 0.028 |
| MA2 | Stars | 5.2 | 0.064 | 0.71 | 0.045 | 5.2 | 0.065 | 0.70 | 0.045 |
| | ISM | 6.0 | 0.058 | 0.56 | 0.032 | 5.0 | 0.055 | 0.69 | 0.038 |
| MA3 | Sun | 9.6 | 0.029 | 0.54 | 0.016 | 8.9 | 0.030 | 0.54 | 0.017 |
| | Solar System | 8.2 | 0.033 | 0.58 | 0.019 | 5.3 | 0.040 | 0.63 | 0.025 |
| MA4 | High energy | 3.8 | 0.083 | 0.51 | 0.042 | 3.8 | 0.096 | 0.56 | 0.054 |
| | Relativity | 4.0 | 0.058 | 0.38 | 0.022 | 3.0 | 0.093 | 0.40 | 0.037 |
| MA5 | Instrumentation | 4.5 | 0.062 | 0.83 | 0.052 | 5.4 | 0.066 | 0.80 | 0.052 |

Table 7 and Table 8 list the top ten keywords according to impact for Italy and INAF, respectively.

**Table 7. Top ten keywords of Italian astronomy (2010-2012)**

| Keyword | Total cit | Cit/paper | Int | Rank Italy | Italy/World | INAF/World | INAF/Italy | University/Italy |
|---|---|---|---|---|---|---|---|---|
| GRB | 14190 | 11.6 | 3.07 | 2 | 0.095 | 0.048 | 0.51 | 0.37 |
| Stars - clusters | 19231 | 10.1 | 2.28 | 3 | 0.095 | 0.056 | 0.59 | 0.39 |
| Stars - abundances | 9987 | 11.6 | 2.86 | 3 | 0.095 | 0.068 | 0.72 | 0.23 |
| High energy | 16794 | 11.5 | 3.99 | 3 | 0.093 | 0.042 | 0.45 | 0.39 |
| Cosmic rays | 12657 | 11.7 | 3.27 | 3 | 0.088 | 0.030 | 0.34 | 0.38 |
| Galaxies – active | 22156 | 11.9 | 3.34 | 3 | 0.087 | 0.055 | 0.63 | 0.33 |
| Radiative processes | 15410 | 11.4 | 3.45 | 4 | 0.086 | 0.044 | 0.51 | 0.39 |
| gamma-ray | 82907 | 11.1 | 3.16 | 4 | 0.083 | 0.041 | 0.49 | 0.38 |
| AGN | 24280 | 12.2 | 3.17 | 4 | 0.083 | 0.048 | 0.58 | 0.38 |
| Galaxies – clusters | 23297 | 13.4 | 2.84 | 3 | 0.080 | 0.045 | 0.56 | 0.26 |

Italy is second (behind United States) in Gamma Ray Bursts. It is third (behind United States and Germany) in High Energy, Stars: clusters, Stars: abundances, Cosmic rays, Galaxies: active, Galaxies: clusters, Stars: populations, Galaxies: halos. Its impact is over 9% of world one for the first four of these keywords.

**Table 8 Top ten keywords of INAF (2010-2012)**

| Keyword | Total cit | Cit/ paper | Int | Rank Italy | Italy/ World | INAF/ World | INAF/ Italy | University/ Italy |
|---|---|---|---|---|---|---|---|---|
| Stars - abundances | 9987 | 11.6 | 2.86 | 3 | 0.095 | 0.068 | 0.72 | 0.23 |
| Stars - clusters | 19231 | 10.1 | 2.28 | 3 | 0.095 | 0.056 | 0.59 | 0.39 |
| Galaxies – active | 22156 | 11.9 | 3.34 | 3 | 0.087 | 0.055 | 0.63 | 0.33 |
| Stars - populations | 20007 | 13.8 | 2.87 | 3 | 0.080 | 0.054 | 0.68 | 0.28 |
| GRB | 14190 | 11.6 | 3.07 | 2 | 0.095 | 0.048 | 0.51 | 0.37 |
| AGN | 24280 | 12.2 | 3.17 | 4 | 0.083 | 0.048 | 0.58 | 0.38 |
| Stars - atmosphere | 9518 | 13.0 | 2.16 | 6 | 0.057 | 0.047 | 0.82 | 0.17 |
| Supernovae | 42139 | 15.5 | 2.58 | 4 | 0.072 | 0.046 | 0.64 | 0.23 |
| Galaxies - individual | 13152 | 8.9 | 2.81 | 5 | 0.060 | 0.046 | 0.77 | 0.20 |
| Galaxies – clusters | 23297 | 13.4 | 2.84 | 3 | 0.080 | 0.045 | 0.56 | 0.26 |

In general, there is a good correlation between the impact of Italian astronomy in the different topics and their internationalization factor (see Figure 3). Two exceptions (fields with high internationalization and weak Italian presence) are cosmic background radiation and Sun: structure.

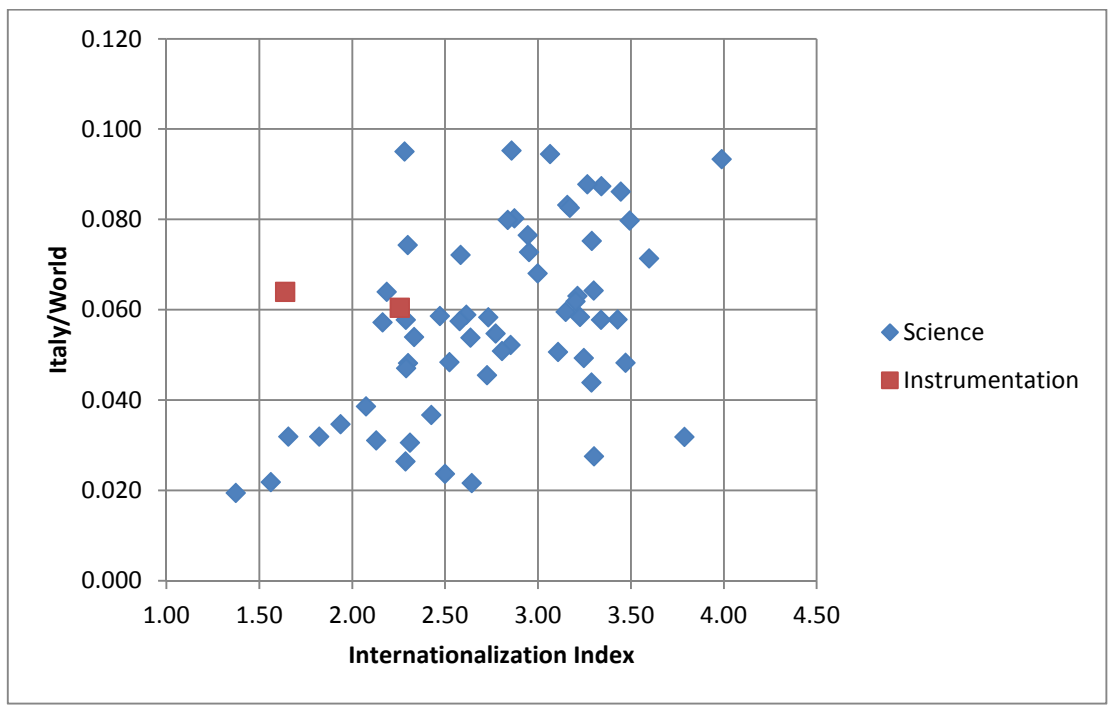

**Figure 3. Correlation between the impact of Italian astronomy and the internationalization index of the various keywords.**

There is a very good correlation between the impact of Universities and of the research Institutes (INAF+other research institutes) (see Figure 4). The research institutes are strong almost only in those fields where also universities are strong. The two most deviating cases are Stars: formation and Stars: Abundances and Composition; in both cases, these fields are much stronger at INAF than expected from their strength in the University.

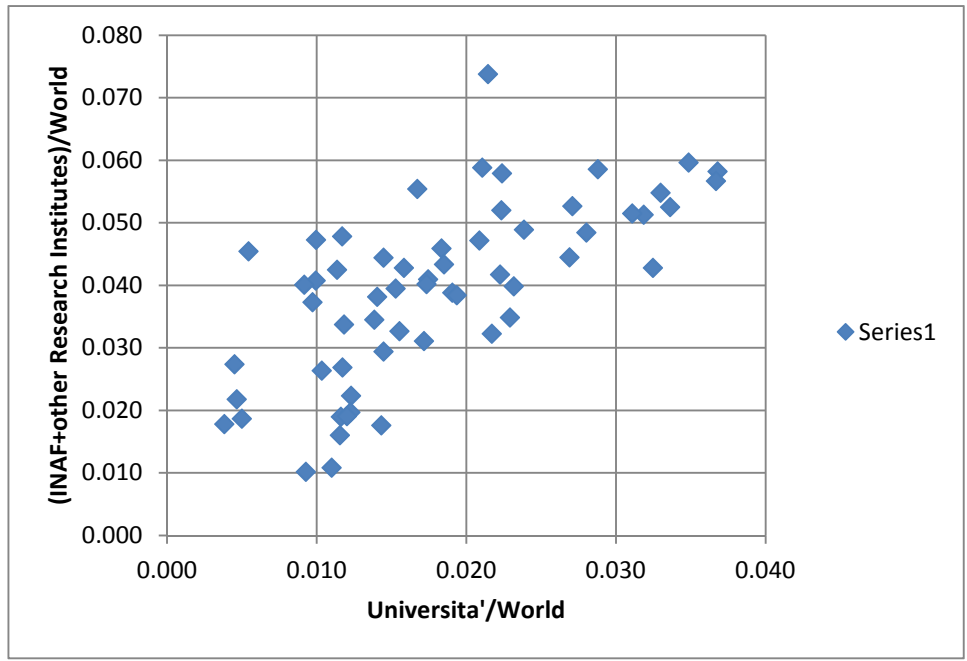

Figure 4. Correlation between the impact of Universities and of research Institutes

Figure 5 compares the impact of Italian astronomy for individual keywords for the periods 2005-2007 and 2010-2012. As expected, there is a good overall correlation: those fields where Italian astronomy was strong 7 years ago still are the fields where it is strong now. However, there is some loss on the ability to be leader in a field: there were 4 keywords where the impact of Italian astronomy was >10% in 2005-2007, and none today.

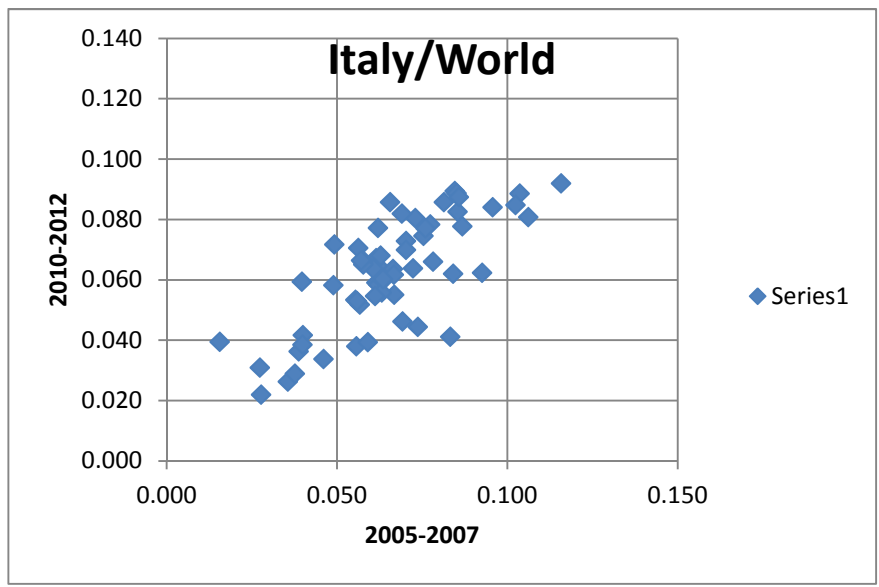

Figure 5. Comparison between the impact of Italian astronomy for individual keywords for the periods 2005-2007 and 2010-2012

## Fashion Index

We may also define a *Fashion-Index* as the ratio between citations for papers published in the period 2010-2012 divided for the same quantity for the period 2005-2007. This *Fashion Index* should be taken with care, because it might reflect seasons of great success for a particular field rather than long-term trends. Figure 6 compare the Impact of Italian astronomy with this Fashion Index. There is no obvious correlation.

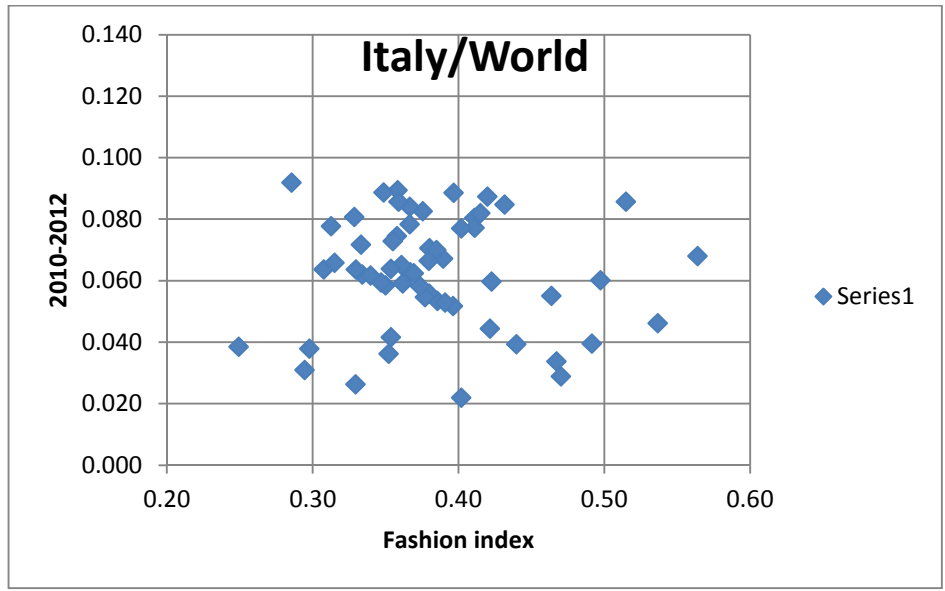

Figure 6. Impact of Italian astronomy versus Fashion Index for individual keywords

Table 9. Impact of Italian astronomy in the ten most fashionable keywords

| Keyword | World citations | Fashion Index | 2010-2012 | | 2005-2007 | |
|---|---|---|---|---|---|---|
| | | | Rank Italy | Italy/World | Rank Italy | Italy/World |
| Galaxies: high redshifts | 20345 | 0.56 | 5 | 0.068 | 5 | 0.063 |
| Hydrodynamics | 27813 | 0.54 | 6 | 0.046 | 3 | 0.069 |
| Cosmic Rays | 12657 | 0.51 | 3 | 0.086 | 6 | 0.066 |
| Dark Matter | 52680 | 0.50 | 5 | 0.060 | 5 | 0.064 |
| Sun: structure | 5085 | 0.49 | 11 | 0.040 | 11 | 0.016 |
| Extrasolar planets | 30577 | 0.47 | 10 | 0.029 | 8 | 0.038 |
| Radiative transfer | 10820 | 0.47 | 7 | 0.034 | 7 | 0.046 |
| Galaxies: star formation | 14023 | 0.46 | 4 | 0.067 | 5 | 0.067 |
| Early Universe | 21277 | 0.44 | 6 | 0.039 | 5 | 0.059 |
| Synchrotron, Compton, Radiative processes | 15410 | 0.43 | 4 | 0.085 | 5 | 0.102 |
| *Weighted Average* | | | *5.9* | *0.054* | *5.5* | *0.061* |

Table 9 gives the impact of Italian astronomy in the ten most fashionable keywords. Italian astronomy has good results for Galaxies: high redshifts and Cosmic Rays and in part for Synchrotron, Compton, Radiative processes, but on average has a lower impact in these fields than over the whole astronomy, and for most of these fields Italy has lost ground in the last years. This unsatisfactory result is mainly driven by the poor result in Extrasolar planets.

10# Ranking by individual keywords

## General keywords

**Table 10. General keywords**

| Keyword | Total cit | Cit/paper | Int | Rank Italy | Italy/World | INAF/World | INAF/Italy | University/Italy |
|---|---|---|---|---|---|---|---|---|
| Surveys | 82742 | 14.1 | 3.30 | 5 | 0.064 | 0.043 | 0.67 | 0.29 |
| Catalogs | 24016 | 13.1 | 3.49 | 5 | 0.080 | 0.049 | 0.61 | 0.34 |
| Gravitation | 61958 | 11.0 | 2.30 | 6 | 0.048 | 0.019 | 0.39 | 0.32 |
| Hydrodynamics | 27813 | 14.6 | 2.07 | 6 | 0.039 | 0.020 | 0.50 | 0.30 |
| Gravitational lensing | 10530 | 11.6 | 3.29 | 7 | 0.044 | 0.017 | 0.39 | 0.33 |
| Radiative Transfer | 10820 | 10.2 | 2.29 | 7 | 0.026 | 0.020 | 0.78 | 0.18 |
| Radio | 36915 | 10.0 | 3.60 | 4 | 0.071 | 0.039 | 0.55 | 0.38 |
| Infrared | 60891 | 12.1 | 3.43 | 7 | 0.058 | 0.037 | 0.64 | 0.33 |

## Macroarea 1: Galaxies and Cosmology

**Table 11. Role of Italian astronomy in different areas according to selected keywords – Macroarea 1 Galaxies and cosmology**

| Keyword | Total cit | Cit/paper | Int | Rank Italy | Italy/World | INAF/World | INAF/Italy | University/Italy |
|---|---|---|---|---|---|---|---|---|
| Galaxies - starbursts | 10455 | 15.5 | 3.21 | 6 | 0.063 | 0.038 | 0.61 | 0.37 |
| Galaxies – star formation | 14023 | 14.1 | 3.47 | 4 | 0.048 | 0.030 | 0.62 | 0.36 |
| Galaxies – active | 22156 | 11.9 | 3.34 | 3 | 0.087 | 0.055 | 0.63 | 0.33 |
| Galaxies – clusters | 23297 | 13.4 | 2.84 | 3 | 0.080 | 0.045 | 0.56 | 0.26 |
| Galaxies – formation | 28510 | 18.5 | 2.77 | 6 | 0.055 | 0.038 | 0.69 | 0.28 |
| Galaxies – high redshift | 20345 | 19.4 | 3.23 | 5 | 0.058 | 0.039 | 0.68 | 0.30 |
| Galaxies – evolution | 42242 | 15.9 | 3.20 | 5 | 0.062 | 0.042 | 0.68 | 0.30 |
| Galaxies - nucleus | 25515 | 11.6 | 2.95 | 4 | 0.073 | 0.045 | 0.61 | 0.33 |
| Galaxies - individual | 13152 | 8.9 | 3.15 | 5 | 0.060 | 0.046 | 0.77 | 0.20 |
| Galaxies – structure | 12200 | 13.1 | 3.11 | 6 | 0.051 | 0.040 | 0.78 | 0.20 |
| Galaxies – kinematics | 12385 | 12.1 | 2.52 | 4 | 0.048 | 0.031 | 0.65 | 0.29 |
| Galaxies – halo | 14495 | 15.1 | 2.47 | 3 | 0.059 | 0.035 | 0.59 | 0.27 |
| Cosmology | 81045 | 15.4 | 2.58 | 5 | 0.058 | 0.029 | 0.50 | 0.30 |
| Dark ages | 31506 | 10.2 | 2.29 | 6 | 0.047 | 0.035 | 0.75 | 0.21 |
| Large scale structure | 19407 | 13.6 | 2.64 | 5 | 0.054 | 0.032 | 0.59 | 0.21 |
| Cosmological parameters | 11943 | 20.2 | 3.25 | 5 | 0.049 | 0.029 | 0.60 | 0.19 |
| Cosmic background | 10871 | 29.9 | 3.30 | 5 | 0.028 | 0.011 | 0.40 | 0.42 |
| Dark matter | 52680 | 17.7 | 2.33 | 5 | 0.054 | 0.019 | 0.35 | 0.40 |
| Dark energy | 20863 | 17.5 | 2.43 | 8 | 0.037 | 0.019 | 0.51 | 0.28 |
| Early universe | 21227 | 15.5 | 2.13 | 6 | 0.031 | 0.007 | 0.24 | 0.39 |

## Macroarea 2: Stars and Interstellar Matter

Table 12. Role of Italian astronomy in different areas according to selected keywords – Macroarea 2 Stars and Interstellar Matter

| Keyword | Total cit | Cit/ paper | Int | Rank Italy | Italy/ World | INAF/ World | INAF/ Italy | University/ Italy |
|---|---|---|---|---|---|---|---|---|
| Extrasolar planets | 30577 | 13.3 | 2.64 | 10 | 0.022 | 0.017 | 0.78 | 0.18 |
| Stars – evolution | 12373 | 10.9 | 2.62 | 5 | 0.059 | 0.040 | 0.68 | 0.25 |
| Stars – formation | 15451 | 9.8 | 2.81 | 6 | 0.051 | 0.044 | 0.87 | 0.11 |
| Stars - populations | 20007 | 13.8 | 2.87 | 3 | 0.080 | 0.054 | 0.68 | 0.28 |
| Stars - abundances | 9987 | 11.6 | 2.86 | 3 | 0.095 | 0.068 | 0.72 | 0.23 |
| Stars - clusters | 19231 | 10.1 | 2.28 | 3 | 0.095 | 0.056 | 0.59 | 0.39 |
| Stars - atmosphere | 9518 | 13.0 | 2.16 | 6 | 0.057 | 0.047 | 0.82 | 0.17 |
| Supernovae | 42139 | 15.5 | 2.58 | 4 | 0.072 | 0.046 | 0.64 | 0.23 |
| Interstellar Matter | 11929 | 8.9 | 3.34 | 6 | 0.058 | 0.032 | 0.56 | 0.34 |

## Macroarea 3: Sun and Solar System

Table 13. Role of Italian astronomy in different areas according to selected keywords – Macroarea 3: Sun and Solar System

| Keyword | Total cit | Cit/ paper | Int | Rank Italy | Italy/ World | INAF/ World | INAF/ Italy | University/ Italy |
|---|---|---|---|---|---|---|---|---|
| Sun - atmosphere | 14838 | 5.6 | 1.66 | 8 | 0.032 | 0.016 | 0.50 | 0.45 |
| Sun - dynamo | 6420 | 4.4 | 1.37 | 12 | 0.019 | 0.010 | 0.51 | 0.48 |
| Sun – structure | 5085 | 8.8 | 3.79 | 11 | 0.032 | 0.023 | 0.73 | 0.14 |
| Sun – planet interactions | 11107 | 4.1 | 1.56 | 7 | 0.022 | 0.007 | 0.31 | 0.50 |
| Solar system planets | 7896 | 6.3 | 1.94 | 8 | 0.035 | 0.019 | 0.53 | 0.36 |
| Comets | 4436 | 4.3 | 2.31 | 9 | 0.031 | 0.019 | 0.61 | 0.38 |
| Minor bodies | 8661 | 3.5 | 1.82 | 8 | 0.032 | 0.020 | 0.61 | 0.38 |

## Macroarea 4: High Energy and Relativity

Table 14. Role of Italian astronomy in different areas according to selected keywords – Macroarea 4: High Energy and Relativity

| Keyword | Total cit | Cit/ paper | Int | Rank Italy | Italy/ World | INAF/ World | INAF/ Italy | University/ Italy |
|---|---|---|---|---|---|---|---|---|
| High energy | 16794 | 11.5 | 3.99 | 3 | 0.093 | 0.042 | 0.45 | 0.39 |
| X-ray | 52717 | 10.4 | 2.95 | 4 | 0.076 | 0.041 | 0.54 | 0.37 |
| gamma-ray | 82907 | 11.1 | 3.16 | 4 | 0.083 | 0.041 | 0.49 | 0.38 |
| pulsar | 12088 | 9.4 | 3.29 | 4 | 0.075 | 0.034 | 0.45 | 0.43 |
| GRB | 14190 | 11.6 | 3.07 | 2 | 0.095 | 0.048 | 0.51 | 0.37 |
| AGN | 24280 | 12.2 | 3.17 | 4 | 0.083 | 0.048 | 0.58 | 0.38 |
| X-ray binaries | 8513 | 7.4 | 2.30 | 5 | 0.074 | 0.047 | 0.63 | 0.30 |
| Radiative processes | 15410 | 11.4 | 3.45 | 4 | 0.086 | 0.044 | 0.51 | 0.39 |
| Cosmic rays | 12657 | 11.7 | 3.27 | 3 | 0.088 | 0.030 | 0.34 | 0.38 |
| Relativity | 27166 | 10.5 | 2.29 | 4 | 0.058 | 0.022 | 0.38 | 0.40 |
| Astroparticles | 44795 | 10.8 | 2.19 | 4 | 0.064 | 0.022 | 0.35 | 0.35 |

## Macroarea 5: Instrumentation

**Table 15. Role of Italian astronomy in different areas according to selected keywords – Macroarea 5. Instrumentation**

| Keyword | Total cit | Cit/paper | Int | Rank Italy | Italy/World | INAF/World | INAF/Italy | University/Italy |
|---|---|---|---|---|---|---|---|---|
| SPIE – papers | 4351 | 0.8 | 1.64 | 4 | 0.064 | 0.054 | 0.84 | 0.13 |
| SPIE - citations | 3624 |  | 2.26 | 5 | 0.061 | 0.050 | 0.82 | 0.13 |

## Distribution of h-factor

Another interesting parameter to be considered is the distribution of the h-factor for astronomers that are members of the INAF's scientific Macroareas (this parameter is not meaningful for those people mainly involved in Instrumentation). We gathered data for a total of 506 investigators from ADS. The distribution of the h-factor is given in Figure 7. The average value is 26.8 and the median value is 25. There are 43 investigators with an h-factor above 50 that is considered excellence according to topitalianscientists.org/Top_italian_scientists_VIA-Academy.aspx .

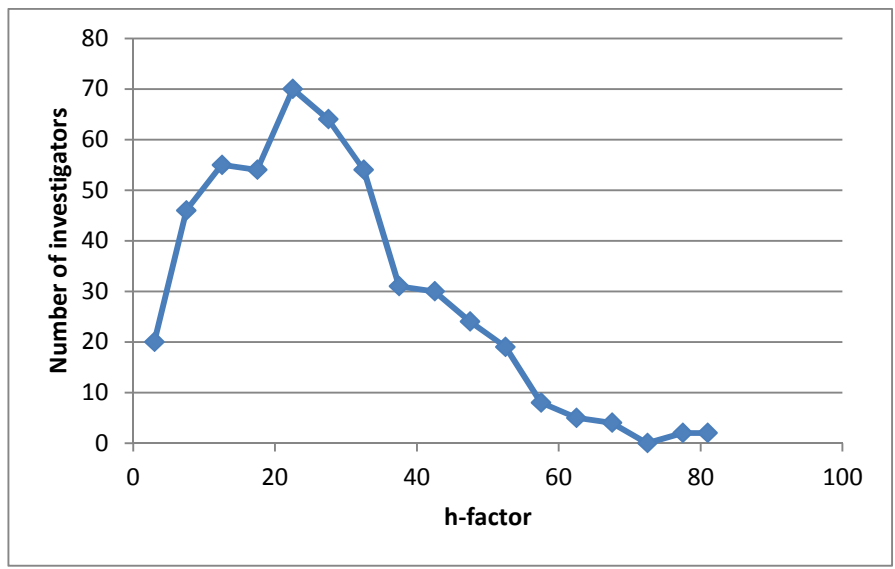

**Figure 7. Distribution of h-factor for scientist members of INAF Macroareas 1 to 4.**

We may compare these values with the statistics given by Abt (2012), who gives the following mean h-indexes of IAU members: France = 21.1, Germany = 24.2, UK =23.5, USA = 24.5. The values we obtain for INAF staff are slightly higher, but this difference might be due to the time lapse between the two studies. A survey over about 50 Italian astronomers showed that on average they doubled their citations over the last 5 years. This implies a ~20% increase in 1.3 year. Since h-factor depends on the square root of citations, I expect 10% increase of the h-factor in 1.3 yr. I then obtain values of 24.4 (average) and 22.7 (median) as appropriate for the same epoch considered by Abt.

I conclude that the values for Italian astronomers are very similar to those for the other four major astronomical countries.

## Acknowledgments

I wish to thank Stephen S. Murray and Alberto Accomazzi for help using the ADS database.

## Appendix

### Notes on individual keywords

*Sun: atmosphere* = "Sun: atmosphere" OR "Sun: chromosphere" OR "Sun: corona" OR "Sun: coronal mass ejections (CMEs)" OR "Sun: faculae" OR "Sun: plages" OR "Sun: filaments" OR "Sun: prominences" OR "Sun: flares" OR "Sun: granulation" OR "Sun: sunspots" OR "Sun: x-ray" OR "Sun: gamma-ray" OR "Sun: transition region"

*Sun: dynamo* = "Sun: dynamo" OR "Sun: activity"

*Sun: structure* = "Sun: helioseismology" OR "Sun: oscillations" OR "Sun: fundamental parameters" OR "Sun: interior"

*Sun/planet interactions* = "Interplanetary magnetic field" OR "planets: Ionospheres" OR "planets: Magnetospheres" OR "planets: Exosphere" OR "Solar wind" OR "solar-terrestrial relations" OR "Sun: heliosphere"

*Solar system planets* = "Planets" – ("Planets" AND "Stars")

*Extrasolar planets* = "Planets" AND "Stars"

*Minor bodies* = "Minor planets" OR "Asteroids" OR "Meteorites"

*Stars: formation* = "Stars: formation" – ("Stars: formation" AND "Galaxies")

*Stars: abundances* = "Stars: abundances" OR "Stars: composition"

*Stars: clusters* = "Globular clusters" OR "Open clusters and associations" OR "Star clusters"

*Stars: atmospheres* = "Stars: atmospheres" OR "Stars: photospheres" OR "Stars: chromospheres" OR Stars: coronae"

*Interstellar matter* = "Interstellar matter" OR "Nebulae"

*Galaxies kinematics* = "Galaxies: kinematics" OR "Galaxies: dynamics"

*Dark ages* = "Dark ages" OR "First stars" OR "Reionization"

*GRB* = "GRB" OR "gamma-ray burst"

*AGN* = "AGN" OR 'Active galactic nuclei"

*Radiative processes* = "Synchrotron" OR "Compton" OR "Radiative processes"